\documentstyle[12pt]{article}
\documentstyle{fleqn}

\begin{document}
\begin{flushright}
OU-HET 241 (revised)
\\
March,   1996 \\
\end{flushright}
\vskip2cm
\begin{center}
{\bf  THE RUBAKOV-CALLAN EFFECT AND BLACK HOLES
\footnote{
To appear in Proceedings of the Workshop ``
{\it Frontiers in Quantum Field Theory} " in honor of the
 60th birthday of Professor Keiji Kikkawa
}
}
\end{center}
\vskip0.5cm
\begin{center}
{\large 
T.  KUBOTA
}

\vskip0.5cm
{\it
Department of Physics, Osaka University

Toyonaka, Osaka, 560,  Japan
}
\end{center}
\vskip0.5cm
\small
\begin{quotation}
\noindent
The Rubakov-Callan effect is reexamined by 
considering the gravitational effects caused by the 
heavy monopole mass. Assuming that the Higgs vacuum 
expectation value is as large as (or larger than) the 
Planck mass, we show  that the calculational scheme 
of Rubakov and Callan may be extended in the presence 
of curved background field. It is argued that the 
density of the fermion condensate around a magnetically 
charged black hole is modified in an intricate way.

\end{quotation}
\normalsize
\vskip0.7cm
\noindent
{\bf 1}\hskip0.5cm{\bf  Introduction}

\vskip0.2cm
\noindent
It was early in spring of 1983 when  I moved to  Osaka 
University that I began to have the  good fortune to 
work with Professor Keiji Kikkawa as one of his 
research associates. No sooner than I arrived at Osaka,  
Kikkawa suggested to me to work with him on the finite 
temperature effect on the monopole catalysis of the 
baryon number violation,  then a brand-new  phenomenon 
predicted by Rubakov [1] and Callan [2]. We started to 
work on this fresh  idea together with  H.S. Song 
and the paper [3] is an outcome of our joint efforts.

The Rubakov-Callan effect attracted  a lot of broad 
attention because of its phenomenological and astrophysical 
importance.  In my opinion, however, it is even more 
 attractive  because of the  novelty and boldness 
shown at  the cutting edge of the development of 
non-perturbative field theories. Here I resume  
the Rubakov-Callan effect, but  as a 
theoretical laboratory with a hope of gaining an  
insight into  gravitational effects in quantum 
field theories.  

Rubakov and Callan have shown that the fermion condensate
around a magnetic monopole is long-ranged, i.e., 
falling off  by an inverse power of the 
distance from the monopole.  Kikkawa et al. [3] on the 
other hand  have claimed  that the
condensate at finite temperature becomes short-ranged, and 
that the typical  length scale is set  by the inverse of 
temperature.
If we consider the gravitational effect caused by the 
heavy monopole mass, we have another 
length scale, namely, the Planck length. It looks more 
than likely that the spatial dependence of the fermion 
condensate should be at variance with Rubakov-Callan's 
 at this length scale. I  will argue  in this short note 
that this is in fact the case, and will give necessary   
calculational formulae.
\vskip1cm
\noindent
{\bf 2}\hskip0.5cm{\bf  Magnetically Charged Black Holes}

\vskip0.2cm
\noindent
To make things as clear as possible, we consider the 
conventional Einstein gravity coupled with $SU(2)$ 
gauge theory which spontaneously breaks down to $U(1)$ 
group by an $SU(2)$-triplet Higgs scalar $\Phi ^{a}$. 
We are concerned with the behavior of left-handed 
fermion $\Psi $ around a  magnetic monopole in the 
presence of the gravitational background. 
The action consists of three parts
\begin{equation}
S=
-\frac{1}{16\pi G}
\int d^{4}x\sqrt{-g}R+
S_{{\rm boson }}+S_{{\rm fermion}},
\end{equation}
where the bosonic and fermionic actions are given, 
respectively, by
\begin{equation}
S_{{\rm boson}}=\int d^{4}x \sqrt{-g}
\left \{-\frac{1}{4}(F_{\mu \nu}^{a})^{2}+
\frac{1}{2}\vert D_{\mu}\Phi ^{a}\vert^{2}-\frac{\lambda}{2}
\left (\vert \Phi ^{a}\vert ^{2}-v^{2}\right )^{2}\right \},
\end{equation}
\begin{equation}
S_{{\rm fermion}}=i\int d^{4}x\sqrt{-g}
{\bar \Psi }e_{\ell}^{\mu}\gamma ^{\ell}
\left (
\partial _{\mu}-\frac{i}{4}\omega ^{mn}_{\mu}\sigma _{mn}
-\frac{i}{2}eA_{\mu}^{a}\tau ^{a}
\right )\frac{1-\gamma _{5}}{2}\Psi .
\end{equation}
Here the gauge coupling is denoted by $e$.
We will consider the static and spherically symmetric 
metric 
\begin{equation}
ds^{2}=B(r)dt^{2}-A(r)dr^{2}-r^{2}(d\theta ^{2}+
\sin ^{2}\theta d\phi ^{2})
\end{equation}
as a solution to the Einstein equation 
and $e^{\mu} _{\ell}$ and $\omega _{\mu}^{mn}$ in Eq. (3) 
are the vierbein and spin connections respectively 
associated with the above metric. 

The 'tHooft-Polyakov monopole in the gravitational 
background (4) has been considered in literatures [4,5,6]. 
Numerical solutions to the Einstein-Yang-Mills-Higgs
 equations are also available. 
There are two mass scales in this system: one is the 
Higgs vacuum expectation value $v$, and the other is 
the Planck mass $M_{P}$. If $v$ is as large as $M_{P}$ 
and the Schwarzschild radius is comparable to the 
monopole size, then the monopole should become a black
 hole.  Lee, Nair and  Weinberg [6] classified the 
classical soultions, drawing a ``phase diagram"
 according to the value of $v$ and the monopole mass.   
They also discussed the relation of their solutions to 
the Reissner-Nordstr{\" o}m solution. 

We are now interested in solving  the monopole-fermion 
dynamics, considering (strong) quantum fluctuations 
of Yang-Mills field in the presence of the gravitational 
classical background (4). Following Rubakov and Callan, 
we will consider a particular type of quantum 
fluctuations around the 'tHooft-Polyakov monopole 
configuration,  
\begin{eqnarray}
\Phi ^{a}\tau ^{a}&=&vh(r)\tau ^{1},
                         \\
A_{\mu}^{a}\tau ^{a}e_{0}^{\mu}&=&
\frac{2}{e}a_{0}(r, t)\tau ^{1},
\hskip1cm 
A_{\mu}^{a}\tau ^{a}e_{1}^{\mu}=
\frac{2}{e}a_{1}(r, t)\tau ^{1},
                   \nonumber  \\
A_{\mu}^{a}\tau ^{a}e_{2}^{\mu}&=&
\frac{u(r)}{er}\tau ^{2},
\hskip2cm 
A_{\mu}^{a}\tau ^{a}e_{3}^{\mu}=
\frac{u(r)}{er}\tau ^{3}+\frac{\cot \theta}{er}\tau ^{1}.
\end{eqnarray}
Here $a_{0}(r, t)$ and $a_{1}(r, t)$ are the quantum 
fluctuations and $u(r)$ and $h(r)$ are the classical part.

To put our system into a solvable form, we have to restrict 
ourselves to a limiting case, in which the radius of the
 monopole   is vanishingly small. More specifically we 
will assume that  $v$ is much larger than $M_{P}$. 
To take  this limiting case implies  $h(r)=1$ and 
$u(r)=0$ outside the horizon. 
The Higgs and gauge fields take the same 
configurations as those  at large distance even at the
 horizon. The bosonic action (2) then becomes 
\begin{equation}
S_{{\rm boson}}=\frac{8\pi }{e^{2}} \int dt  dr 
\sqrt{A(r)B(r)}\;\:r^{2}\: \:  \left \{
\frac{1}{\sqrt{B(r)}}\frac{\partial 
a_{1}(r, t)}{\partial t}-\frac{1}{\sqrt{A(r)}}
\frac{\partial a_{0}(r, t)}{\partial r}\right \}^{2}
\end{equation}
up to an additive constant.
This is an  action analogous to the two-dimensional QED in 
the gravitational background and $a_{0}(r, t)$ and 
$a_{1}(r, t)$ are the vector potential.
\vskip1cm
\noindent
{\bf 3}\hskip0.5cm{\bf  The Fermion Dynamics }

\vskip0.2cm
\noindent
A fermion moving around the monopole comes close to 
the core of the monopole if it is in $S$-wave.
 From here on we will consider only the 
$S$-wave fermions which is defined, if we use the 
chiral basis for the gamma matrices,  by
\begin{equation}
\sqrt{r^{2}\sin \theta \sqrt{B(r)}}
\:\:\Psi ^{T}=( 0, \chi),\hskip1cm
 \chi _{\alpha \ell }(r, t)=\delta _{\alpha \ell}\chi _{1}
(r, t)-i(\tau ^{1})_{\alpha \ell}\chi _{2}(r,t).
\end{equation}
Here $\alpha$  and $\ell$ are spin and isospin indices, 
respectively. The $S$-wave part of the fermionic action 
is reminiscent of the two-dimensional fermion with 
two components if we set
$\chi ^{T}=(\chi _{1}, \chi _{2})$, 
\begin{equation}
S_{{\rm fermion}}=-4\pi \int dt  dr \sqrt{A(r)}\;
\chi ^{\dag}D_{J=0}\; \chi ,
\end{equation}
where the Dirac operator in this case becomes
\begin{equation}
D_{J=0}= -i \left (\frac{1}{\sqrt{B(r)}}\frac{\partial }
{\partial t}+i\tau ^{2}a_{0}(r, t) \right )
+i\tau ^{2}\left (\frac{1}{\sqrt{A(r)}}\frac{\partial }
{\partial r}-i\tau ^{2}a_{1}(r, t)\right ).
\end{equation}

The fermionic action (9) and the  bosonic one (7) are 
saying that our dynamical system looks similar to the 
Schwinger model. There is, however, an important difference, 
that is, the boundary condition to be imposed on the 
fermion field. In the case of Rubakov and Callan, the 
boundary condition was imposed at $r=0$, so that the 
approximation of vanishing monopole size is justified. 
In our case we have assumed  $v\gg M_{P}$ and the 
Schwarzschild radius is greater than the radius of 
 the monopole core. It is the most 
reasonable choice of the boundary condition that the 
flow of the probability should cease to go inside at the
 horizon (located at $r=r_{H}$) so that 
the hermiticity of the Hamiltonian 
is preserved. This is achieved by setting
\begin{equation}
(\tau ^{1}+i\tau ^{2})\chi (r_{H}, t)=0.  
\end{equation}

Now we will take the gauge fixing condition 
$A^{a}_{0}=0$ and will discard $a_{0}(r, t)$.
The  technique used by Rubakov and Callan to solve 
the monopole-fermion dynamics was the bosonization method, 
which is also vital in our case. Let us denote the 
bosonized field  by $\varphi  (r, t)$. The 
fermionic kinetic term 
is expressed by the bosonic one,  
 and  furthermore $\varphi (r, t)$ is 
related to the gauge field via
\begin{equation}
\Box \varphi  (r, t)=\frac{1}{\sqrt{B(r)}}
\frac{\partial a_{1}(r, t)}{\partial t},
\end{equation}
where $\Box $ is the two dimensional Laplacian associated with 
$ds^{2}=B(r)dt^{2}-A(r)dr^{2}$.

A close look at Eqs. (7) and (12) shows that the kinetic 
term of the gauge field may be easily expressed by 
$\varphi  (r, t)$ and we arrive at the effective action
\begin{equation}
S_{{\rm eff}}=S_{{\rm boson}}+S_{{\rm fermion}}=
\frac{1}{2}\int dt dr \sqrt{A(r)B(r)}\varphi  (r, t)
L_{rt}\varphi  (r, t),
\end{equation}
where
\begin{equation}
L_{rt}=\frac{8\pi}{e^{2}}\frac{1}{\sqrt{A(r)B(r)}}
\Box \sqrt{A(r)B(r)}r^{2} \Box +\frac{N_{f}}
{\pi}\Box .
\end{equation}
The number of fermion flavors is denoted by 
$N_{f}$. 
Thus in the curved background as well, the two kinds of 
dynamical degrees of freedom due to fermionic and 
bosonic parts are described by a single scalar field 
$\varphi  (r, t)$.  The boundary condition (11) is 
transcribed into 
$\partial _{r}\varphi  =0$ at $r=r_{H}$.
\vskip1cm
\noindent
{\bf 4}\hskip0.5cm{\bf Gravitational Effects on the 
Fermion Condensate  }

\vskip0.2cm
\noindent
We are now in a position to see how the effect of the 
curved  background appears in particular on the fermion 
condensate around a magnetic monopole. We will consider 
two flavor fermion doublets ($N_{f}=2$). The fermion 
condensate that we are concerned 
with is $<f(r, t)>$, where 
\begin{equation}
f(r, t)=\chi _{1}^{(1)}(r, t)\chi _{1}^{(2)}(r, t)+
\chi _{2}^{(1)}(r, t)\chi _{2}^{(2)}(r, t).
\end{equation}
The superscript denotes the flavor index.

The dynamics described by (13) can be solved if the full 
Green's function $(L_{rt})^{-1}$ is known exactly. 
The operator (14) is very much simplified if we restrict our 
considerations to the Reissner-Nordstr{\" o}m case , 
i.e., $A(r)B(r)=1$. We will denote this Green's function 
by ${\cal P}(r, t; r', t')$ for this particular case 
and the boundary condition is $\partial _{r}{\cal P}=0$
 at the horizon $r=r_{H}$.  
According to the analysis by  Rubakov and Callan, 
the density of the fermion condensate is given by 
$<f(r,t)>\sim \exp \{2{\cal P}(r,t;r,t)\}$, the
 coincidence limit of this Green's function.
 It should be noticed that the  Green's function 
${\cal P}(r,t;r',t')$ is expressed as a combination 
of two  other types of Green's functions defined by 
\begin{eqnarray}
& &\Box {\cal D}(r,t;r',t')=\delta (r-r')\delta (t-t'),
             \nonumber \\
& &\left (\Box +\frac{\kappa}{r^{2}}\right ){\cal R}_
{\kappa}(r,t;r',t')=\delta (r-r')\delta (t-t').
\end{eqnarray}
Here we used the notation 
$\kappa =N_{f}e^{2}/8\pi ^{2}$.

It is rather straightforward to obtain 
${\cal D}(r,t;r',t')$
by using the method of DeWitt and Schwinger [7].
Applying their method to our case, one finds that the 
Green's function is  expressed by the Hankel function 
$H_{0}^{(2)}(\mu \sqrt{-2\sigma})$ and its derivatives.
 Here $\sigma =\sigma (r,t;r',t')$ is the two-dimensional 
geodesic interval between $(r,t)$ and $(r',t')$ 
and $\mu $ is an 
infrared cutoff. The leading term in the DeWitt-Schwinger 
expansion turns out to be 
\begin{equation}
\frac{\Delta}{4\pi}\log [-2\mu ^{2}\sigma (r,t;r',t')],
\end{equation}
where $\Delta $ is the bi-scalar function defined by 
DeWitt. The boundary condition is made satisfied by putting 
mirror images at $2r_{H}-r$ and $2r_{H}-r'$.

The coincidence limit of the Green's function 
${\cal D}(r,t;r,t)$  contributes 
partly to ${\cal P}(r,t;r,t)$, and it  entails 
 the Coulomb interaction between $(r,t)$ and its mirror 
image. The density of the fermion condensate is described 
thus by the bi-scalar functions $\Delta$ and $\sigma $.

As to the other Green's function 
${\cal R}_{\kappa }(r,t;r',t')$ , 
the calculational method of DeWitt [7] 
does not apply in its original form. 
It is, however, easy to see that the term $\kappa /r^{2}$ 
 in (16) is playing the role of infrared regulator 
(or the position-dependent mass), 
and that there must exist a term like $\log r$ in the 
coincidence limit 
${\cal R}_{\kappa}(r,t;r,t)$, 
 contributing to the fermion condensate.

\vskip1cm
\noindent
{\bf 5}\hskip0.5cm{\bf  Summary}

\vskip0.2cm
\noindent
In the present paper we have seen that the mathematical 
framework of Rubakov and Callan may be generalized in an 
analogous way   to the case 
of the curved background. The density of the fermion 
condensation is deformed not simply by the deformation 
of the space-time but also intricately by  the mirror image.
Throughout the present note   the gravity has been 
treated classically and quantum aspects 
of gravity are all neglected. 
It would be of particular interest if we could include 
strong gravitational fluctuations in this framework.

Gregory and Harvey [8] studied the zero-energy soultion 
of the Dirac equation in the presence of the magnetic
 monopole and gravitational background. They argued 
 the possibility of baryon number violating scattering 
processes off the magnetically charged black hole.  
The relation between  their approach and the present 
one will be discussed elsewhere.

\vskip1cm
\noindent
{\bf Acknowledgements}

\vskip0.2cm
\noindent
I would like to express my  sincere thanks to  
Professor Keiji Kikkawa for his continuous     
 guidance and encouragement. The present essay 
 is dedicated to him  on the occasion of his   
sixtieth birthday.  This work is supported in  
part by the Grant in Aid for Scientific        
Research from the Ministry of Education,       
Science and Culture (Grant Number: 06640396).

\vskip1cm
\noindent 
{\bf References}
\begin{description}
\item{[1]}
V.A. Rubakov, Nucl. Phys. {\bf B 203} (1982) 311.
\item{[2]}
C.G. Callan, Phys. Rev. {\bf D 25} (1982) 2141; 
{\it ibidem} {\bf D 26} (1982) 2058.
\item{[3]}
K. Kikkawa, T. Kubota and H.S. Song, Prog. 
Theor. Phys. {\bf 71} (1984) 1346. 
\item{[4]}
F.A. Bais and R.J. Russel, Phys. Rev. {\bf D 11} 
(1975) 2692; Y.M. Cho and P.G.O. Freund, Phys. Rev. 
{\bf D 12} (1975) 1588.
\item{[5]}
P. van Nieuwenhuizen, D. Wilkinson and M.J. Perry, 
Phys. Rev. {\bf D 13} (1976) 778.
\item{[6]}
K. Lee, V.P. Nair and E.J. Weinberg, Phys. Rev. 
{\bf D 45} (1992) 2751; Phys. Rev. Lett. {\bf 68} (1992)
1100.
\item{[7]}
B.S. DeWitt, {\it Dynamical Theory of Groups and 
Fields}  
(Gordon and Breach, New York, 1965); 
J. Schwinger, Phys. Rev. {\bf 82} (1951) 664.
\item{[8]}
R. Gregory and J.A. Harvey, Phys. Rev. {\bf D 46} 
(1992) 3302.
\end{description}
\vskip1cm

\pagebreak
\noindent
{\bf Note added:}

\noindent
After submitting the present paper  for publication, 
I was informed of the related work by Piljin Yi [Phys. 
Rev. {\bf D 49} (1994) 5295]. He also worked out the 
generalization of the Rubakov-Callan's bosonized effective 
action around general static spherically symmetric black holes. 
I would like to thank Dr. Piljin Yi for calling my attention 
to his paper. 

\end{document}